\documentstyle[11pt,epsfig]{article}
\newcommand{\beq}{\begin{equation}}
\newcommand{\eeq}{\end{equation}}
\newcommand\bea{\begin{eqnarray}}
\newcommand\eea{\end{eqnarray}}
\newcommand{\C}{{\cal C}}
\newcommand{\dimn}{\mbox{dim}}

\newcommand{\wb}{b_{2}}
\newcommand{\tr}{\mbox{Tr}}
\newcommand{\cover}{\to}

\newcommand{\E}{{\cal E}}

\makeatletter
\newcommand{\ps@preprint}{%
        \renewcommand{\@oddhead}{\hfil\small{CGPG-99/3-4}}}%
\makeatother
\begin{document}
\title{Quantum causal histories}
\author{Fotini Markopoulou\thanks{fotini@phys.psu.edu}\\
	Center for Gravitational Physics and Geometry\\
	Department of Physics\\
	The Pennsylvania State University\\
	University Park, PA, USA 16802}
\date{March 25, 1999}
\maketitle
\thispagestyle{preprint}  
\vfill
\begin{abstract}
Quantum causal histories are defined to be causal sets with Hilbert spaces
attached to each event and local unitary evolution operators.  The 
reflexivity, antisymmetry, and transitivity properties of a 
causal set are preserved in the quantum history as conditions on the evolution
operators.  A quantum causal history in which transitivity holds can be treated as
``directed'' topological quantum field theory. 
Two examples of such histories are described.   
\end{abstract}
\vfill
\newpage

\section{Introduction}

In general relativity, with compact rather than asymptotically flat boundary conditions,
physical observations are made inside the system that they describe. 
In quantum theory, the observable quantities are meaningful outside 
the system that they refer to.  It is very likely that quantum gravity must
be a ``quantum mechanical relativistic theory". That is, 
a theory where the 
observables can be given as self-adjoint operators on a Hilbert space but which are 
meaningful inside the system that they describe.  

A rough description of what such a theory involves is the following.  
Observations made inside the system are 
closely related to causality in the sense that an inside
observer necessarily splits the history of the system 
into the part that is in the future, the part that is in the past, and --- assuming 
finite speed of propagation of information --- elsewhere.   We may 
call such observables ``internal observables'', characterised by the 
requirement that they refer to information that an observer at a 
point, or a connected region of spacetime, may be able to gain about 
their causal past.  In a previous paper \cite{fm3}, we found that 
these observables can be described by functors from the partially 
ordered set of events in the spacetime to the category of sets.  Such 
a functor codes the relationship between the causal structure and the 
information available to an observer inside the spacetime.  This has 
non-trivial consequences and, in particular, the observables algebra 
is modified even at the classical level.  Internal observables satisfy 
a Heyting algebra, which is a weak version of the Boolean algebra of 
ordinary observables.  This is still a distributional algebra. Namely, 
for propositions $P, Q$ and $ R$, if $P\vee Q$ denotes ``$P$ or $Q$'', and 
$P\wedge Q$ means ``$P$ and $Q$'', then $P\vee(Q\wedge R)= (P\vee 
Q)\wedge(P\vee R)$.  

On the other hand,  quantum mechanics is linear,
and as a result of the superposition principle, 
quantum mechanical propositions are not distributive.
If $P,Q$ and $R$ are projection operators,  
$P\vee(Q\wedge R)$ is not equal to $(P\vee Q)\wedge(P\vee R)$.
Having both quantum mechanics and internal observables in the same theory 
requires finding propositions that have a non-distributive 
quantum mechanical aspect and a distributive causal aspect. 
The aim of the present paper is to define the histories  in which such 
observables may be encountered.  

We will, therefore,  define quantum causal histories, which are
histories that are both quantum mechanical and causal. 
Assuming that a discrete causal ordering (a causal set)
is a sufficient description of the fundamental past/future ordering
needed to qualify observations as being inside the system, 
we find that a quantum causal history
can be constructed by attaching finite-dimensional 
Hilbert spaces to the events of the causal set.    It is then natural 
to consider tensor products of the Hilbert spaces on events
that are spacelike separated.  We define quantum histories with local 
unitary evolution maps between such sets of spacelike separated events. 
The conditions of reflexivity, antisymmetry and transitivity that hold 
for the causal set have analogs in the quantum history which are conditions on the 
evolution operators.   We find that transitivity is a strong physical condition on
the evolution operators and, if imposed, implies that the quantum
causal histories are invariant under directed coarse graining.  

If the the causal set represents the universe, quantum causal histories constitute
a quantum cosmological theory.  Its main notable feature is that there is a Hilbert
space for each event but not one for the entire universe.  Hence, no wavefunction 
of the universe arises.  A consistent intepretation of quantum causal histories and
observatons inside the 
quantum universe can be provided and will appear in a forthcoming paper \cite{fmls4}.

In more detail, 
the outline of this paper is the following.  In section 2 we review causal 
set histories and provide a list of definitions of structures that can 
be found in a causal history and which are used in the quantum causal 
histories.  In particular, we concentrate on acausal sets, sets
of causally unrelated events.  In section 3, we 
introduce the poset of acausal sets, equipped with the appropriate
ordering relation. The definition of quantum causal histories is based on this poset 
and is given in section 4.  The properties of the resulting histories 
are discussed in section 5.  The ordering of a causal set is 
reflexive, antisymmetric and transitive, conditions which are also imposed
on the quantum histories.  The consequences of these properties are 
analyzed in section 6. In particular, we find that transitivity leads to 
directed coarse-graining invariance.  
Two classes of quantum causal histories are 
given as examples in section 7.  Up to this point, the causal histories require a 
choice of a causal set.  In section 8, we remove this restriction and 
provide a sum-over-histories version of quantum causal evolution. 
The quantum causal histories presented here are consistent,
but not all physically meaningful questions can be asked.   There are 
several possibilities for generalisations, some of which 
we outline in the Conclusions.

\section{Causal set histories}

A (discrete) causal history is a causal set of events that
carry extra structure. For example, the causal histories
that were examined in \cite{fmls1, fm2, fmls2} had as
events vector spaces spanned by $SU_q(2)$ intertwiners.
In two dimensions, an exact model of such a causal history has been 
proposed by Ambjorn and Loll \cite{AL} and its continuum limit 
properties have been investigated in \cite{AL,ANRL}.
The dynamics of a 3-dimensional causal spin network history model 
is addressed by Borissov and Gupta in \cite{BG}\footnote{ 
For pure causal set theories (with no extra structure on the events), 
we may note recent work by  Rideout and Sorkin who derived a 
family of stochastic sequential growth dynamics 
for a causal set, with 
very interesting consequences about the classical limit of pure 
causal sets \cite{RS}.  Further, the dynamics of a toy model causal set 
using a suitable quantum measure was proposed by 
Criscuolo and Waelbroek in 
\cite{CW}.}.

In this section, we review the definition of a causal set and 
provide several derivative definitions which will be used in the 
rest of this paper. 

A causal set $\C$ is a partially ordered set whose elements are interpeted
as the events in a history (see \cite{blms,sorkin,dm}). 
We denote the events by $p,q,r,\ldots$.
If, say, $p$ precedes $q$, we write $p\leq q$. The equal option is used when 
$p$ coincides with $q$.   We write $p R q$ when either $p\leq q$ or 
$p\geq q$ holds. 

The causal relation is reflexive, i.e.\ $p\leq p$ for any event $p$.
It is also transitive, i.e.\ if $p\leq q$ and 
$q\leq r$, then $p\leq r$. To ensure that $\C$ has no closed timelike 
loops, we make the causal relation antisymmetric, that is, 
if $p\leq q$ and $q\leq p$, then $p=q$. 
Finally, we limit ourselves to histories with a finite number of events. 

Given a causal set, there are several secondary structures that we
can construct from it and which come in useful in this paper. 
We therefore list them here (see Figures 1 and 2):
\begin{itemize}
\item
The {\it causal past} of some event $p$ is the set of all events 
$r\in\C$ with $r\leq p$. We denote the causal past of $p$ 
by $P(p)$.  
\item
The {\it causal future} of $p$ is the set of all events $q\in\C$ with 
$p\leq q$. We denote it by $F(p)$.
\item
An {\it acausal set}, denoted $a,b,c,\ldots$,  is a set of events in $\C$ that are 
all causally unrelated to each other. 
\item
The acausal set $a$ is a {\it complete past} of the event $p$ 
when every event in the causal past $P(p)$ of $p$ is related to some
event in $a$. It is not possible to add an event from 
$(P(p)-a)$ to $a$ and 
produce a new acausal set. 
\item
Similarly, an acausal set $b$ is a {\it complete future} of $p$
when every event in the causal future $F(p)$ of $p$ is
related to some event in $b$.
\item
A {\it maximal antichain} in the causal set $\C$ is an acausal set $A$ such that 
every event in $(\C-A)$ is causally related to some event in $A$.
\end{itemize}

Similar definitions as the above of past, future, complete past,
and complete future for a single event can be given for acausal sets:
\begin{itemize}
\item
The causal past of $a$ is $P(a)=\bigcup_i P(q_i)$ for all $q_i\in a$.
Similarly, the causal future of $a$ is $F(a)=\bigcup_i F(q_i)$
for all $q_i\in a$.
\item
An acausal set $a$ is a complete past of the acausal set $b$ if every event
in $P(b)$ is related to some event in $a$. 
\item
An acausal set $c$ is a complete future of $b$ if every 
event in $F(b)$ is related to some event in $c$. 
\end{itemize}
Furthermore, 
\begin{itemize}
\item
Two acausal sets $a$ and $b$ are a {\it complete pair}
when $a$ is a complete past of $b$ and $b$ is a complete future
of $a$.  
\item
Two acausal sets $a$ and $b$ are a {\it full pair} when they 
are a complete pair and every event in $a$ is related to every 
event in $b$. 
\item
Two acausal sets $a$ and $b$ {\it cross} when some of the 
events in $a$ are in the future of $b$ and some are in its past. 
\end{itemize}

\begin{figure}
\centerline{\mbox{\epsfig{file=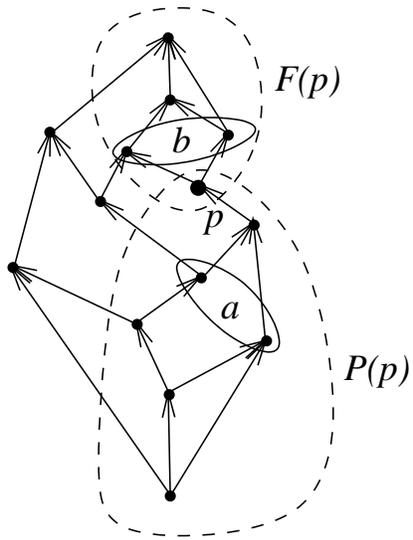}}}
\caption{This is a very small causal history. $P(p)$ is the causal 
past of the event $p$ while $F(p)$ is its causal future.  
The acausal set $a$ is a complete past 
for $p$, and $b$ is a complete future.  } 
\label{PastFuture}
\end{figure}

\begin{figure}
\centerline{\mbox{\epsfig{file=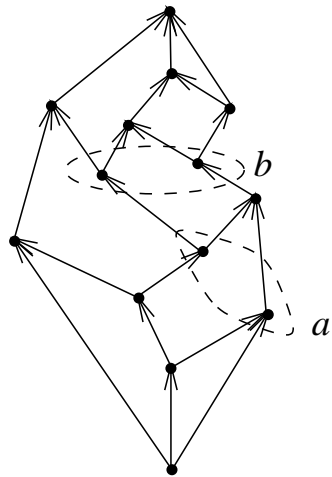}}}
\caption{The acausal sets $a$ and $b$ are a complete pair.} 
\label{completepair}
\end{figure}

\section{The poset of acausal sets}

The set of acausal sets within a given causal set $\C$ is a partially 
ordered set if we define the relation $a\preceq b$ to mean that $a$ is 
a complete past of $b$ and $b$ is a complete future of $a$.  
Reflecting the properties of the underlying causal set, the relation
$\preceq$ is reflexive, transitive and antisymmetric. Let us call this
poset ${\bf A}$. 

It is on this poset of acausal sets that we will base
the quantum version of the causal histories. 
Its properties, therefore, are important constraints 
on the corresponding quantum history.
The main property of ${\bf A}$ that characterises the kind of
quantum theory we will obtain in this paper is that, given acausal sets
$a, b$ and $c$, the following holds ($R$ means either
$\preceq$ or $\succeq$):
\beq
	\mbox{If }aRc, bRc,\mbox{ and }a, b\mbox{ do not cross, then } aRb.
\eeq
That is, given some acausal set $c\in{\bf A}$, all the acausal sets 
that are related to $c$ are also related to each other --- except if they
happen to cross.  
If $a$ and $b$ are ``too close'' to each other
they may cross and then cannot be related by 
$\preceq$. This means that for the chosen $c$ there is not a unique 
complete pair sequence.  
In selecting one of the possible sequences we need to make repeated 
choices of which of two crossing acausal sets to keep. 

\section{Quantum causal histories}
\label{qch}

We will now construct the quantum version, $Q{\bf A}$,  of 
the poset ${\bf A}$. 
We will regard an event $q$ in the causal set as a Planck-scale quantum 
``event'' with a Hilbert space $H(q)$ that stores its possible 
states.   
We require that $H(q)$ is finite-dimensional, which is consistent 
with the requirement that our causal sets be finite. 

Choose an acausal set $a=\{q_1,q_2,\ldots,q_n\}$ in ${\bf A}$.
Since all $q_i\in a$ are causally unrelated to each other, 
standard quantum mechanics dictates that the Hilbert space
of $a$ is
\beq
	H(a)=\bigotimes_{i=1}^n H(q_i).
\eeq
That is, we have a tensor product Hilbert space in $Q{\bf A}$
for each acausal set in ${\bf A}$.

When two acausal sets are related,
$a\preceq b$, there needs to be an evolution operator between
the corresponding Hilbert spaces:
\beq
	E_{ab}:H(a)\longrightarrow H(b).
\label{eq:E}
\eeq
We will impose one more condition on the causal histories that will 
make their present treatment
simpler. We will only consider posets ${\bf A}$ with the
following property:
\beq
	\mbox{ If }a\preceq b,\qquad\dimn\ H(a)=\dimn\ H(b).
\eeq
This restriction is particularly convenient since it allows us to simply 
regard $H(a)$ and $H(b)$
as isomorphic and require that $E_{ab}$ is a unitary evolution operator.  

The poset of acausal sets ${\bf A}$ is reflexive, transitive and antisymmetric. 
We would like to maintain these properties of the causal ordering in the quantum 
theory as analogous conditions on the evolution operators. 
(In other words, we want the quantum causal history to be a functor from 
the poset ${\bf A}$ to the category of Hilbert spaces.)
The analogue of reflexivity is the existence of an operator 
$E_{aa}={\bf 1}_a:H(a)\rightarrow H(a)$
for every acausal set $a$.  $E_{aa}$ has to be the identity
because any other operator from $H(a)$ to itself  would
have to be a new event.
Transitivity in ${\bf A}$ implies that
\beq
	E_{bc} E_{ab}=E_{ac}
\label{eq:transitivity}
\eeq
in $Q{\bf A}$. 
We will return to transitivity and its implications for ${\bf A}$ and
$Q{\bf A}$ in section 6.  

At each event $q$, there is an algebra of observables,
the operators on $H(q)$.
An observable $\widehat{O}_a$ at $a$ becomes an observable 
$\widehat{O}_b$ at $b$ by
\beq
	\widehat{O}_b={E_{ab}}\widehat{O}_a E_{ab}^\dagger.
\eeq

This completes the definition of the causal 
quantum  histories we are concerned with. 
In the next section we discuss the evolution of states which is
allowed in such histories. Then, in section 6, we will come back to the 
imposition of transitivity on the 
evolution operators, a strong condition that dictates the 
form of the resulting histories and their
quantum cosmology interpretation.

\section{Quantum evolution in $Q{\bf A}$}

In this section we discuss the consequences of the definitions of quantum  
histories given above. 
  
\subsection{Products of complete pair sequences}
Evolution maps between complete pairs that are themselves causally 
unrelated to each other may be composed in the standard way, like tensors. That is, consider
complete pairs $a\preceq b$ and $c\preceq d$, with $a$ and $b$ unrelated to 
$c$ or $d$.  Then construct the acausal sets $a\cup c$ and $b\cup d$, 
which form a new complete pair: $(a\cup c)\preceq (b\cup d)$. The evolution 
operator on the composites,
\beq
	E_{(a\cup c)(b\cup d)}:H(a)\otimes H(c)
		\longrightarrow H(b)\otimes H(d),
\eeq
is the product of the operators on the two pairs,
\beq
	E_{(a\cup c)(b\cup d)}=E_{ab}\otimes E_{cd}.
\eeq

\subsection{Projection operators}

The causal structure of ${\bf A}$ means that a projection operator on 
$H(a)$ propagates to the future 
of $a$ in the following way.  
A projection operator 
\beq
	P_a:H(a)\rightarrow V(a)
\eeq
that reduces $H(a)$ to a subspace $V(a)$, can be extended to a 
larger acausal set $a\cup c$. On $H(a)\otimes H(c)$, it is
the new projection operator 
\beq
	P_{a\cup c}=P_a\otimes{\bf 1}_c
		:H(a)\otimes H(c)\longrightarrow V(a)\otimes H(c).
\eeq
By using the evolution operator  
$E_{(a\cup c)(b\cup d)}=E_{ab}\otimes E_{cd}$ on the enlarged 
projection operator we obtain a projection operator $P_{b}\otimes
1_{d}$ on the future acausal set $b\cup d$.

\subsection{Evolution in ${\bf A}$ can be independent of ${\C}$}
\label{rho_problem}

Consider a complete pair $a\preceq b$ in which $a=a_1\cup a_2$ and
$b=b_1\cup b_2$. The corresponding Hilbert spaces are
\beq
	H(a)=H(a_1)\otimes H(a_2)\qquad\mbox{and}
	\qquad H(b)=H(b_1)\otimes H(b_2),
\eeq
and $E_{ab}$ is the evolution operator that corresponds to the causal relation 
$a\preceq b$. 

Choose some state $|\psi\rangle\in H(a_1)$ by 
acting on $H(a_1)$ with the projection operator 
$|\psi\rangle\langle\psi|$.  This implies that, in $H(a)$, 
we have chosen the state $|\psi\rangle\otimes|\psi_{a_2}\rangle$,
for some $|\psi_{a_2}\rangle\in H(a_2)$, using the projection 
operator $(|\psi\rangle\langle\psi|)\otimes{\bf 1}_{a_2}$.
We can use $E_{ab}$ on this state to 
obtain the state
\beq
	|\psi_b\rangle=E_{ab}\left(|\psi\rangle\otimes|\psi_{a_2}\rangle
			\right)
\eeq
in $H(b)$. 
 
If, for any reason, we need to 
restrict our attention to $b_2$, we can  
trace over $b_1$ to find that the original state 
$|\psi\rangle\in H(a_2)$ gives rise to the density matrix:
\bea
	\rho_\psi(b_2)&=&\tr_{b_1}|\psi_b\rangle\langle\psi_b\|\\
		&=&\tr_{b_1}\left[E_{ab}\left(|\psi\rangle
		\langle\psi|\otimes{\bf 1}_{a_2}\right)E_{ab}^\dagger\right].
\label{eq:rho}
\eea
At this point, the following question arises.  What if $a_1$ is not 
in the causal past of $b_2$ and, for example, we have these 
causal relations:
\beq
	\begin{array}{c}\mbox{\epsfig{file=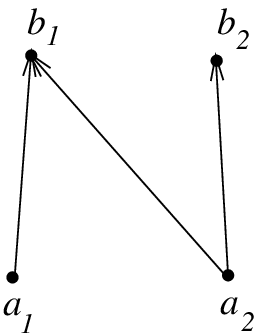}}\end{array}
\nonumber
\eeq
Does the evolution we defined on $Q{\bf A}$ violate causality in the 
underlying causal set $\C$?
  
This question illuminates several features of the quantum causal 
histories.  The very first thing to note is that we get the same 
acausal poset ${\bf A}$ for many different causal sets.  The
operator $E_{ab}$ refers to ${\bf A}$ and does not distiguish 
the different possible underlying causal sets.  

There is a simple solution to the above apparent embarassment.  
Instead of promoting the events of the causal set to Hilbert spaces, 
we may attach the Hilbert spaces to the edges, and the evolution 
operators to the events.  An event in the causal set, then, becomes 
an evolution operator from the tensor product of the Hilbert spaces 
on the edges ingoing to that event, to the tensor product of the 
outgoing ones.  Since the set of ingoing and the set of outgoing edges 
to the same event are a full pair (i.e.\ a complete pair in which all 
events in the past acausal set are related to all the events in the 
future one), the above problem will not arise.  Conceptually, this 
solution agrees with the intuition that events in the causal set 
represent change, and, therefore, in the quantum case they 
should be represented as operators.   In section \ref{ex2}, we discuss 
the example of quantum causal histories with the Hilbert spaces on the edges for 
trivalent causal sets.

\subsection{Propagation by a density matrix requires a complete pair}
\label{aw}
According to (\ref{eq:rho}), given a state $|\psi\rangle\in H(a)$,
we can obtain the density matrix $\rho_\psi(\wb)$ for the acausal set
$\wb$ in the future of $a$.  This uses the fact that
$\wb$ is a subset of an acausal set $b$ that forms a complete 
pair with $a$.  

Consider this configuration:
\beq
	\begin{array}{c}\mbox{\epsfig{file=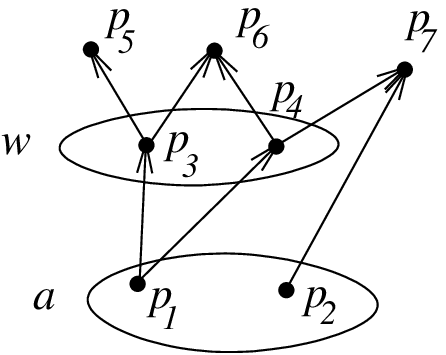}}\end{array}
\nonumber
\label{eq:awset}
\eeq
The initial acausal set is $a=\{p_1,p_2\}$.  The acausal set $w=\{p_3, p_4\}$
is in the future of $a$.
Given $|\psi\rangle\in H(a)$, can we obtain a density matrix on $H(w)$?
The answer is no, since there is no acausal set that contains $w$
and is maximal in the future of $a$. (The problem is the 
$p_2\leq p_7$ relation.)

There are, therefore,  acausal sets in the future of $a$ which cannot be 
reached by the evolution map $E$.  

\section{Directed coarse-graining}

The main idea in this paper is that the quantum version of some causal history
is a collection of Hilbert spaces connected by evolution operators that
respect the structure of the poset ${\bf A}$ we started with.  For this 
reason, in section 4, we imposed reflexivity, transitivity and 
antisymmetry to the operators $E_{ab}$.   

Transitivity is  a strong condition on the quantum history.   
One should keep in mind that it was first imposed on causal sets 
because it holds for the causal structure of Lorentzian spacetimes. 
Significantly, it does not just encode properties of the
ordering of events but also the fact that a 
Lorentzian manifold is a {\it point set}. In general relativity, an event
is a point and this has been imported 
in the causal set approach.  

To analyse this a little further, let
us introduce a notation that indicates when two events $p$ and $q$ are 
related by a shortest causal relation, i.e., no other
event occurs after  $p$ and before $q$. This is the 
{\it covering relation}:
\begin{quote}
$\bullet$ The event $q$ {\it covers} $p$ if $p\le q$ 
and there is no other event $r$ with $p\le r\le q$. We denote this by 
$p\cover q$. 
\end{quote}

The following are worth noting. For a finite causal set, 
transitivity means that the order 
relation determines, and is determined by, the covering relation, since
$p\le q$ is equivalent to a finite sequence of covering relations
$p=p_1\cover p_2 \ldots\cover p_n=q$. On the other hand, 
in the continuum (for example the real line
${\bf R}$) there are no pairs $p,q$ such that $p\cover q$ \cite{DavPri}. 
Hence, in a continuum spacetime, it is simply
not meaningful to consider an ordering that is not transitive.
Non-transitive ordering requires distinguishing between the covering 
relations and the resulting transitive ones.  This distinction is not 
possible in the continuum case. 

In short, for events that are
points, it is sensible to expect that if $p$ leads to $q$ and $q$ to $r$,
then $r$ is in the future of $p$.  If, however, the events were (for
example) spacetime regions of some finite volume, with overlaps, then 
it is unclear whether transitivity would hold. 
(See also section 2.4 in \cite{CJI2}.)
In the causal histories we consider here, an event is a 
Hilbert space. It is, therefore, an open question 
whether it is sensible to impose
reflexivity, transitivity and antisymmetry on the ordering of the 
Hilbert spaces.  We choose to first 
impose them, then find what the implications are, and if they are 
unphysical,  go back and check which of the three conditions 
should not be maintained in a quantum causal ordering.  

On the positive side, there is a very interesting advantage to 
maintaining transitivity.  Using (\ref{eq:transitivity}),
we have the benefit of a {\it directed coarse-graining} invariance of the 
quantum history.  For example, if we are handed 
\beq
	\begin{array}{c}\mbox{\epsfig{file=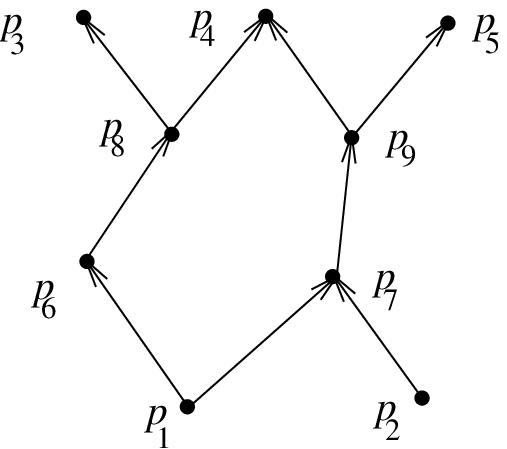}}\end{array}
\nonumber
\eeq
and need to go from $H(p_1)\otimes H(p_2)$ to $H(p_3)\otimes H(p_4)\otimes
H(p_5)$, we can reduce it to
\beq
	\begin{array}{c}\mbox{\epsfig{file=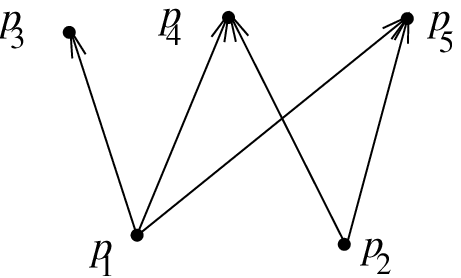}}\end{array}.
\nonumber
\eeq
Clearly, several initial graphs will give the same coarse-grained graph.  
We return to this in section \ref{sums}.

It is very interesting to note that the coarse-graining implied by 
transitivity can be used to improve the 
propagation of density matrices that we discussed in section \ref{aw}.
We can coarse-grain the causal set depicted in (\ref{eq:awset}) by 
considering the events $p_3,p_4,p_5,p_6,p_7$ to be the acausal set 
$\bar{w}$.  This is a coarse-graining in the sense that we ignore any
causal relations between these events. We then obtain:
\beq
	\begin{array}{c}\mbox{\epsfig{file=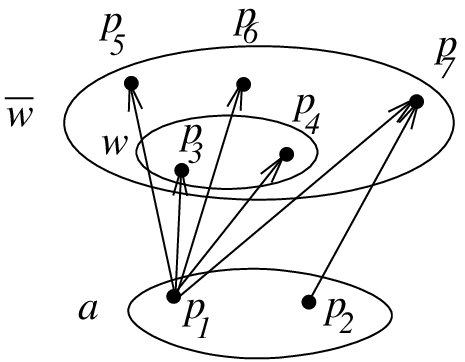}}\end{array}.
\nonumber
\eeq
In this causal set, $a$ and $\bar{w}$ are a complete pair.  It is then 
possible to use an evolution operator $E_{a\bar{w}}$ to take the state 
$|\psi\rangle\in H(a)$ to a state in $H(\bar{w})$ and then trace over 
$(\bar{w}-w)$ to obtain a density matrix on $H(w)$.

\section{Examples}

In this section, we provide two examples of the quantum causal histories we
defined in section \ref{qch}.

\subsection{Discrete Newtonian evolution}
A discrete Newtonian history is a universe with a preferred time and 
foliation.  It is represented by a poset ${\bf A}$ which 
is a single complete pair sequence:
\beq
	\ldots a_n\longrightarrow a_{n+1}\longrightarrow a_{n+2}\ldots .
\eeq
The corresponding quantum history is then
\beq
	\ldots H_n\longrightarrow H_{n+1}\longrightarrow H_{n+2}
	\ldots.
\eeq
with all the Hilbert spaces isomorphic to each other. 

We can denote by $E$ the evolution operator from $H_n$ to $H_{n+1}$.
Evolution from $H_n$ to $H_m$ is then given by 
\beq
	E_{nm}=E^{m-n}.
\eeq

We may compare this universe to the standard one in quantum theory. 
There, there is a single Hilbert space for the entire universe and evolution 
is given by the unitary operator $e^{iH\delta t}$.  In the above, there 
is a sequence of identical and finite-dimensional 
Hilbert spaces.  Since evolution is by discrete steps, 
we may set $\delta t=1$. Then $E=e^{iH}$, for some hermitian operator $H$
and $E_{nm}=e^{i(m-n)H}$.

\subsection{A planar trivalent graph with Hilbert spaces on the edges}
\label{ex2}

This example is a history with multifingered time \cite{fm2}.
It is a planar trivalent graph with finite-dimensional Hilbert spaces 
living on its {\it edges}.  Trivalent means either two ingoing and one outgoing 
edges at a node, or two outgoing and one ingoing.  We exclude nodes with 
no ingoing or no outgoing edges. 

\begin{figure}
\centerline{\mbox{\epsfig{file=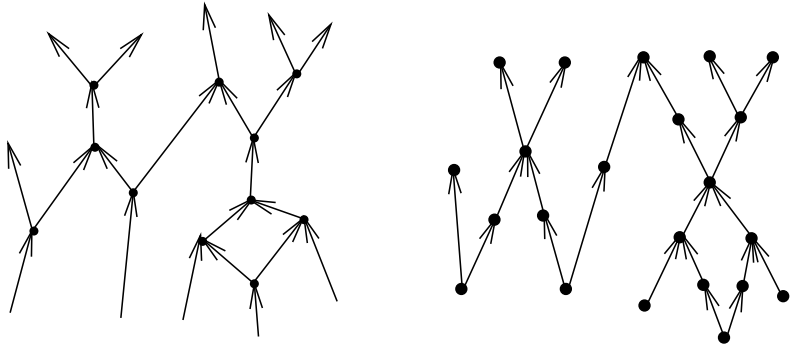}}}
\caption{A causal set and its edge-set.} 
\label{CandEC}
\end{figure}

From a given planar trivalent causal set $\C$, 
we can obtain what we will call its {\it edge-set}, $\E\C$. This is 
a new graph which has the covering relations of $\C$ 
(the edges, not including any transitive ones) as its nodes. 
The covering relations in $\C$ are also ordered and these relations are the 
edges in $\E\C$.  Figure \ref{CandEC} shows an example of a causal set 
and its edge-set.  

We now take the  poset ${\bf A}$ 
of $\E\C$ and construct a quantum history from 
it by assigning Hilbert spaces to the nodes of $\E\C$.  
The very interesting property of $\E\C$ is that 
it can be decomposed into pieces, generating evolution moves that take 
two events to one, or split one event into two. That is,
$\E\C$ can be decomposed to these two full pairs:
\beq
	\begin{array}{c}\mbox{\epsfig{file=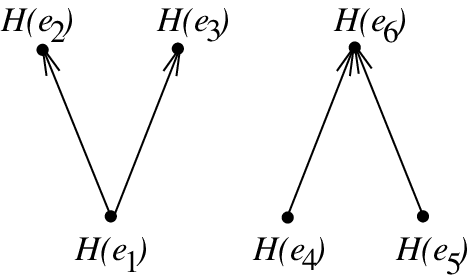}}\end{array}.
\nonumber
\eeq
(This is generally the 
case, not only for trivalent graphs.)  Such a decomposition is not possible 
in some general $\C$ and therefore, 
in view of the problem we encountered in \ref{rho_problem}
above, $\E\C$ provides an advantage over $\C$.

To be able to employ unitary 
evolution operators, we need  $\dimn\ H(e_1)=\dimn\ H(e_2)+\dimn\ H(e_3)$ 
and $\dimn\ H(e_4)+\dimn\ H(e_5)=\dimn\ H(e_6)$.  
One can check that this can be done consistently for all the events in the causal set. 

Although $\C$ is  trivalent, the nodes of $\E\C$ have valence 2,3, or 4. 
The list of possible edges in $\C$ and the corresponding nodes in $\E\C$ is:
\beq
	\begin{array}{c}\mbox{\epsfig{file=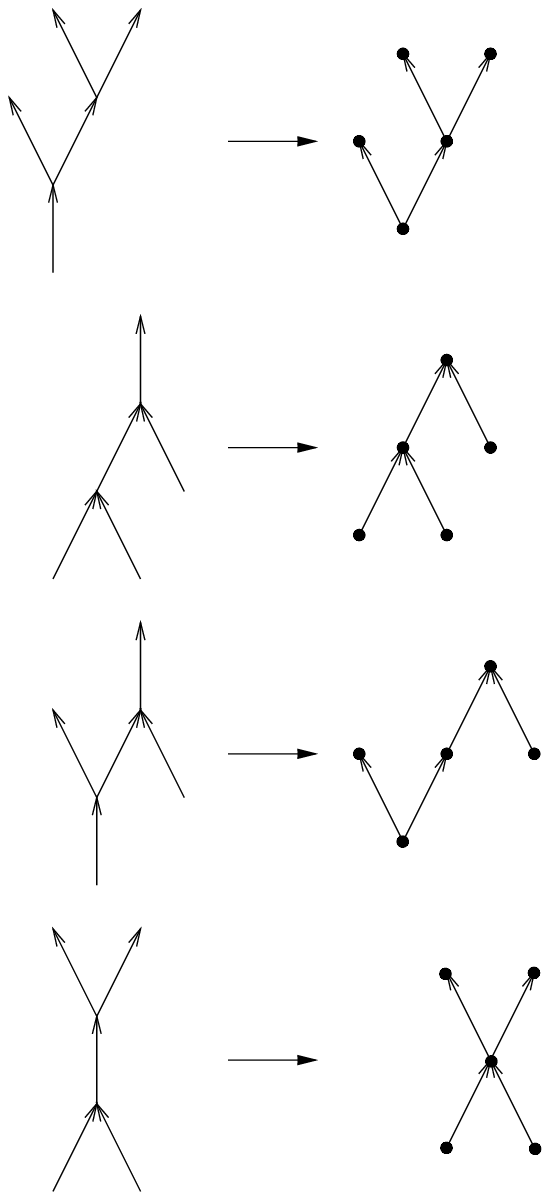}}\end{array}
\nonumber
\eeq
 
There is a substantial interpretational difference between Hilbert spaces on the (covering) 
causal relations and on the events.  A state space on the edge
$p\leq q$
is most naturally interpeted as ``the state space of $p$ as seen by $q$''.
If there are two edges coming 
out of $p$, to $q$ and $r$, then there are two Hilbert spaces in $\E\C$, 
interpeted as the Hilbert space of $p$ as seen by $q$ and the Hilbert space of $p$
as seen by $r$. On the other hand, a Hilbert space placed on the event $p$ is 
absolute, in the sense that is independent of who is observing it. 

\section{Summing over histories}
\label{sums}

The quantum causal histories we discussed in this paper require a
fixed poset ${\bf A}$.  This is also necessary in the treatment of 
observers in a classical causal set universe
in \cite{fm3}.  Since there is no physical reason that some  causal set should 
be preferred over all others, this restriction is unappealing. 
In this  section we outline a ``sum-over-histories'' version of the evolution 
in section \ref{qch}, which also applies to \cite{fm3} and can be used in any 
further work on quantum observers inside the universe. 

An acausal set is a set of points.  We have considered causal sets 
where all events have at least one ingoing and one outgoing edge. 
Then, that $a,b$ are a complete pair $a\preceq b$ means that there is a directed 
graph with the set $a$ as its domain and the set $b$ as its codomain. Let us 
denote this graph by $\gamma(a,b)$. 
A graph with $b$ as its codomain and one with 
$b$ as its domain may be composed. 

When ${\bf A}$ is given, there is one known graph $\gamma(a,b)$ connecting
$a$ and $b$. If ${\bf A}$ is not fixed, we can sum over all graphs 
that we can fit between $a$ and $b$ (that have a finite number
of nodes).  This leads to a sum-over-histories version of the evolution in section 
\ref{qch}.  

Let us call $E_{ab}^\gamma$ the evolution operator (as in equation (\ref{eq:E}))
when the underlying graph is $\gamma(a,b)$.  The transition amplitude from a 
state $|\psi_a\rangle\in H(a)$ to a state $|\psi_b\rangle\in H(b)$ for this 
particular graph is 
\beq
	A_\gamma=\langle\psi_b|E^\gamma_{ab}|\psi_a\rangle.
\eeq
If the graph is not fixed, we may sum over all the possible ones:
\beq
	A_{ab}=\sum_\gamma \langle\psi_b|E^\gamma_{ab}|\psi_a\rangle.
\label{eq:A}
\eeq

Now note that transitivity defines equivalence classes 
of graphs between given acausal sets. 
Here is an example of  a series of graphs that, by transitivity, 
have the same causal relations as far as $a$ and $b$ are concerned. Specifically, 
$p_1,p_2,p_3\leq p_4$ and $p_3\leq p_4,p_5$ in all of these graphs:
\beq
	\begin{array}{c}\mbox{\epsfig{file=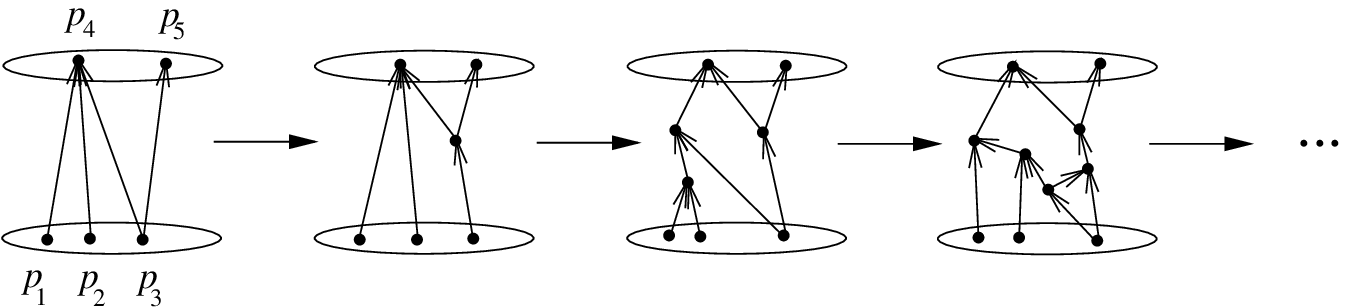}}\end{array}
\nonumber
\eeq
All of the above correspond to the same evolution operator in (\ref{eq:A}). 

It is intriguing to compare this to the triangulation invariance of topological 
quantum field theory.  Transitivity can be intepreted as a {\it directed} triangulation
invariance.  We will return to this in future work.

\section{Conclusions}

We saw that it is possible to promote a causal set to a quantum one by 
taking the events to be finite-dimensional Hilbert spaces. 
It is natural to consider tensor products of the Hilbert spaces on events
that are spacelike separated.  This led us to replace the causal set 
by the poset of acausal sets ${\bf A}$. We defined quantum histories with local 
unitary evolution maps between complete pairs in ${\bf A}$.\footnote{Discrete 
evolution by 
unitary operators is also considered by P.\ Zizzi, in work to appear.}

We explored several of the features of these causal histories.  A property of 
${\bf A}$ is that it splits into distinct sequences of complete pairs.  Consequently, 
this places restrictions on which Hilbert spaces can be reached from a given one. 
The conditions of reflexivity, antisymmetry and transitivity that hold 
for ${\bf A}$ were imposed on the quantum history as conditions on the 
evolution operators.   The most interesting consequences were from transitivity, 
which gave rise to invariance under directed coarse-graining. 
This needs further investigation, since the physical assumption behind 
transitivity is pointlike events.  It will be interesting to consider 
events which are extended objects and find what ordering is suitable in this 
case.

We were able to tensor together the Hilbert spaces for two acausal sets 
when they are spacelike separated and have no events in common.  However, it 
is natural to consider cases where an acausal set is a subset of a larger one.
It appears that, if we need to use acausal sets, we should enrich ${\bf A}$ 
with the inclusion relation, that is, use a poset with {\it two} ordering 
relations, causal ordering and spacelike inclusion. While this is straightforward 
in the plain causal set case, it becomes tricky in the quantum histories. 
For example, we may start from the acausal sets $a=\{q_1,q_2\}$ and 
$a^\prime=\{q_1,q_2,q_3\}$.  In the poset ${\bf A}$ we have $a\subset a^\prime$.
In $Q{\bf A}$, the corresponding state spaces are $H(a)=H(q_1)\otimes H(q_2)$
and $H(a^\prime)=H(q_1)\otimes H(q_2)\otimes H(q_3)$.  However, there 
is no natural way in which $H(a)$ is a subspace of $H(a^\prime)$.
That is, the set inclusion relation is not directly preserved in the quantum 
theory. 

In the quantum histories we discussed, we restricted the past and future 
Hilbert spaces to have the same dimension.  This is the simplest case
and needs to be generalised.  A related issue is
the properties of the individual Hilbert spaces. 
For example, if we employ the causal spin network models of \cite{fmls2}, 
the evolution operators should respect the $SU(2)$ invariance of the 
state spaces.  More generally, a
discrete quantum field theory toy model can be constructed by inserting matter 
field algebras on the events.  This will be addressed in future work. 

Finally, we set up $Q{\bf A}$ having in mind a functor from a poset to Hilbert spaces, 
taking the elements and arrows of the poset into Hilbert spaces and operators 
which preserve the properties of the original poset, i.e.\ reflexivity, antisymmetry 
and transitivity.   It is also possible to use
graphs between sets of events (rather than a fixed causal set), 
as outlined in section \ref{sums}. 
In this case, the quantum causal histories become similar to topological 
quantum field theory except, importantly, they are directed graphs 
(or triangulations). 
The coarse-graining invariance relations can be calculated 
for given fixed valence of the covering relations. 

On the interpretational side, the main thing to note is that the causal history is 
a collection of Hilbert spaces which itself is not a Hilbert space.  
According to quantum theory,  
we can take tensor products of events that are not causally related.  
We cannot, for example, take the tensor 
product of all the Hilbert spaces in the history to be the Hilbert space of the entire 
history.\footnote{Taking tensor products of all Hilbert spaces in the 
history is, however, exactly what is done in the consistent histories approach 
of Isham \cite{CJI}.}  As a result, the causal
quantum cosmology is not described in terms of a wavefunction of 
the universe.  Individual 
events (or observers on the events) can have states and wavefunctions but 
the entire universe does not.  Further discussion of the interpretation of the 
quantum causal histories will appear in \cite{fmls4}. 

\section*{Acknowledgment}

I am grateful to Chris Isham for his comments on transitivity and the 
pointlike structure of spacetime. This paper has benefited from discussions 
with Lee Smolin on quantum theory and the interpetation of such histories 
as quantum cosmologies.  I am also grateful to Sameer Gupta and Eli Hawkins
for discussions on causal histories. 

This work was supported by NSF grants PHY/9514240 and PHY/9423950
to the Pennsylvania State University and a gift from the Jesse 
Phillips Foundation.


\newpage

\begin{thebibliography}{99}
\small{
\bibitem{AL} Ambjorn J and Loll R, 1998,
``Non-perturbative Lorentzian Quantum Gravity, Causality and Topology Change''
{\it  Nucl Phys } {\bf  B536} 407.
%
\bibitem{ANRL} Ambjorn J, Nielsen JL, Rolf J and Loll R, 
1998, ``Euclidean and Lorentzian Quantum Gravity - Lessons from Two Dimensions'',
Chaos Solitons Fractals 10, 177 (hep-th/9806241).
%
\bibitem{blms} Bombelli L, Lee J, Meyer D and 
 R, 1987,
{``Space-time as a causal set''}, {\it Phys Rev Lett}\/ {\bf 59}
521.
%
\bibitem{BG} Borissov R and Gupta S, 1998,
``Propagating spin modes in canonical quantum gravity'', 
Phys.Rev. D60, 024002 (gr-qc/9810024).
%
\bibitem{RS} D P Rideout and R D Sorkin, 1999, ``A Classical 
Sequential Growth Dynamics for Causal Sets'', gr-qc/9904062.
%
\bibitem{CW} Criscuolo A and Waelbroeck H, 1998, ``Causal Set 
Dynamics: A Toy Model'', Class.Quant.Grav. 16, 1817 
(gr-qc/9811088).
%
\bibitem{DavPri} B.A.\ Davey and H.A.\ Priestley, ``Introduction to 
Lattices and Order'', Cambridge University Press, Cambridge, 1992.
%
\bibitem{CJI2} Isham C J, 1994, {`` Quantum Logic and the Histories Approach to 
Quantum Theory''}, {\it J Math Phys} {\bf 35} 2157.
%
\bibitem{CJI} Isham C J, 1997, ``Topos Theory and Consistent 
Histories: The Internal Logic of the Set of all Consistent Sets'', 
{\it Int J Theor Phys} {\bf 36} 785.
%
\bibitem{fm2} Markopoulou F, 1997, {``Dual formulation of spin network
evolution''}, gr-qc/9704013.
%
\bibitem{fm3} Markopoulou F, 1998, {``The internal logic of causal sets:
What the universe looks like from the inside''}, Commun.Math.Phys, to 
appear. 
%
\bibitem{fmls1} Markopoulou F and Smolin L, 1997, {``Causal 
evolution of spin networks''}, {\it Nucl Phys} {\bf B508} 409. 
%
\bibitem{fmls2} Markopoulou F and Smolin L, 1998, {``Quantum 
geometry with intrinsic local causality''}, {\it Phys Rev}\/ D {\bf 58}, 
084032.
%
\bibitem{fmls4} Markopoulou F and Smolin L, 1999, ``Causal quantum cosmology'',
to appear.
%
\bibitem{dm} Meyer D A, 1988, {\sl The Dimension of Causal Sets}, PhD Thesis,
Massachussets Institute of Technology.
%
\bibitem{sorkin} Sorkin R, 1990, {``Space-time and causal sets''} 
in Proc.\ of SILARG VII Conf., Cocoyoc, Mexico.
}
\end{thebibliography}
\end{document}